\documentclass[journal,12pt]{iopart}
\pdfoutput=1
\usepackage[numbers,sort,comma,compress]{natbib}
\usepackage{iopams}
\usepackage{setspace}
\usepackage{amssymb}
\usepackage{CJK}
\usepackage{graphicx}
\usepackage{subfigure}
\usepackage{multirow}
\usepackage{dcolumn}
\usepackage{enumerate}
\usepackage{lineno}
\usepackage{makecell}
\usepackage{booktabs}
\usepackage{gensymb}
\usepackage{verbatim}
\usepackage[cal=rsfs,scr=rsfs]{mathalfa}
\usepackage{caption}
\usepackage{dutchcal}
\usepackage[colorlinks,linkcolor=blue,anchorcolor=blue,citecolor=blue]{hyperref}

\begin{document}
\title[FKS And DL-FPD Enabled Cone-Beam CT Joint Spectral Imaging]{Fast KV-Switching and Dual-Layer Flat-Panel Detector Enabled Cone-Beam CT Joint Spectral Imaging}

\author{Hao Zhou\textsuperscript{1,2}, Li Zhang\textsuperscript{1,2}, Zhilei Wang\textsuperscript{1,2},  and Hewei Gao\textsuperscript{1,2,*}}

\address{1. Department of Engineering Physics, Tsinghua University, Beijing, 100084, China}
\address{2. Key Laboratory of Particle \& Radiation Imaging (Tsinghua University), Ministry of Education, China}
\address{*Author to whom correspondence should be addressed.}
\ead{hwgao@tsinghua.edu.cn}

\vspace{10pt}

\begin{abstract}

	\textbf{Purpose}: Fast kV-switching (FKS) and dual-layer flat-panel detector (DL-FPD) technologies have been actively studied as promising dual-energy solutions for FPD-based cone-beam computed tomography (CBCT). However, CBCT spectral imaging is known to face challenges in obtaining accurate and robust material discrimination performance due to the limited energy separation. To further improve CBCT spectral imaging capability, this work aims to promote a source-detector joint spectral imaging solution which takes advantages of both FKS and DL-FPD, and to conduct a feasibility study on the first tabletop CBCT system with the joint spectral imaging capability developed. 
	
	\textbf{Methods}: A noise performance analysis using the Cramér-Rao lower bound (CRLB) method is conducted. The CRLB for basis material after a projection-domain material decomposition is derived, followed by a set of numerical calculations of CRLBs, for the FKS, the DL-FPD and the joint solution, respectively. In this work, the first FKS and DL-FPD jointly enabled multi-energy tabletop CBCT system, to the best of our knowledge, has been developed in our laboratory. To evaluate its spectral imaging performance, a set of physics experiments are conducted, where the multi-energy and head phantoms are scanned using the 80/105/130kVp switching pairs and projection data are collected using a prototype DL-FPD. To compensate for the slightly angular mismatch between the low- and high-energy projections in FKS, a dual-domain projection completion scheme is implemented. Afterwards material decomposition is carried out by using the maximum-likelihood method, followed by reconstruction of basis material and virtual monochromatic images. 
	
	\textbf{Results}: The numerical simulations show that the joint solution can lead to a significant improvement in energy separation and lower noise levels. The physics experiments confirmed the feasibility and superiority of the joint spectral imaging, whose CNR of the multi-energy phantom were boosted by an average improvement of 21.9\%, 20.4\% for water and 32.8\%, 62.8\% for iodine when compared with that of the FKS and DL-FPD in fan-beam and cone-beam experiments, respectively. 
	
	\textbf{Conclusions}: A feasibility study of the joint spectral imaging for CBCT by utilizing both the FKS and DL-FPD was conducted, with the first tabletop CBCT system having such a capability being developed, which exhibits improved spectral imaging performance as expected.

\end{abstract}

\textbf{Keywords}: fast kV-switching, dual-layer flat-panel detector, joint spectral imaging, material decomposition.

\section{Introduction}
\label{sec:introd}
\par Spectral computed tomography (CT) enables material discrimination by decomposing the energy-dependent linear attenuation coefficients of materials using spectral projections at  different effective energy levels. Compared with the conventional single-energy CT, spectral CT can provide more quantitative information of the scanned object such as the effective atomic number and the electronic density, or alternatively decomposed material components\cite{mccollough2020principles}. Spectral CT can effectively eliminate beam-hardening artifacts caused by the polychromatic spectra used in the single-energy CT, as well as suppress metal artifacts when virtual monochromatic images (VMI) at higher keV are produced. In last decades, spectral CT imaging has grown rapidly in clinical applications such as the gout detection\cite{glazebrook2011identification,bongartz2015dual}, stroke detection\cite{mangesius2021dual,van2021virtual}, and angiography\cite{sun2022pulmonary,harvey2019impacts}. 
\par So far, a variety of spectral imaging technologies have been developed, which can be primarily categorized into the source-based (e.g, dual-kVp by dual-source or kV-switching), the detector-based (e.g, dual-layer or photon counting detector), and the filter-based (e.g, splitting-filter or spectral filter/modulator) methods\cite{mccollough2020principles,deng2022multi,tivnan2022design,cai2023benefits}. In recent years, cone-beam CT (CBCT) spectral imaging has also made significant progress, aiming at improving quantitative performance of CBCT and promoting its various application potentials\cite{gang2012cascaded,cassetta2020fast,muller2016interventional,zhu2022feasibility,cai2023benefits,shi2020characterization,deng2022multi,staahl2021performance,wang2023dual,wang2021high,tivnan2022design,zhu2023super}. Among them, fast kV-switching (FKS) and dual-layer flat-panel detector (DL-FPD) based technologies have already been prototyped or made available in the development mode in the product\cite{cassetta2020fast,wang2021high,shi2020characterization,zhu2023super}. For FKS, low- and high-energy projection data sets are acquired by alternating the tube voltage, with spectral projections quite closely aligned temporally and spatially. It can provide relatively a good energy separation to ensure a relatively stable material decomposition. Due to the shift in rotation in the acquisition of low- and high-energy projection views, interpolation of the sinogram is still necessary when performing material decomposition in the projection domain. To balance the noise of low- and high-energy projection pairs, the exposure time or tube current need to be adjusted wisely in FKS. In the multi-detector CT (MDCT) scanner, it takes less than half a second for a CT scan with tube voltage switching between 80 and 140 kVp in a frequency higher than 2000 Hz\cite{hsieh2009tu}. Fortunately, when applying FKS to CBCT, such a high switching frequency is not necessary due to the limited readout speed of flat-panel detector (FPD). In the literature, the feasibility of FKS for CBCT spectral imaging has been demonstrated\cite{muller2016interventional,cassetta2020fast}. Müller $et$ $al$ measured the stability of FKS in C-arm system directly, confirming it is possible to perform FKS scans on a clinical C-arm CBCT system with some custumed settings in the scanning procedure\cite{muller2016interventional}. Cassetta $et$ $al$ evaluated the dual-energy CT imaging performance using the FKS in the development mode of the on-board imager on a commercial radiotherapy device, where the VMIs and the relative electron density images can provide additional information with potential applications in image guided radiotherapy\cite{cassetta2020fast}. 

\par For DL-FPD, it can acquire two projections that are spectrally distinct but spatially aligned using a single X-ray exposure thanks to the two layers of detector panels. Again, spatially aligned projection data can allow projection-domain material decomposition, leading to an intrinsic reduction in beam hardening artifacts. However, DL-FPD also has some limitations for CBCT spectral imaging, including a moderate energy separation, possibly unpaired signal levels between top and bottom layers. In the literature, the feasibility of DL-FPD for CBCT spectral imaging has also been demonstrated, in terms of dual-energy radiography, high resolution material decomposition and tumor tracking\cite{shi2020characterization,wang2021high,staahl2021performance}. Recently, a hybrid material decomposition was also proposed for head CBCT using DL-FPD, with a feasibility study showing a significant improvement of robustness in material decomposition\cite{wang2023dual}. 

\par In real application, due to the scatter, afterglow, and other non-ideal aspects, CBCT spectral imaging is known to still face challenges in terms of accurate and robust material discrimination and high-quality spectral reconstruction. This is because the energy separation by either FKS or DL-FPD, alone, is still limited, along with apparently unpaired signal levels in the effective low- and high-energy projections in real applications, not to mention the X-ray scatter in cone-beam scan which will make the material decomposition almost impossible if no correction is applied. As a result, a more robust and accurate spectral imaging approach is highly desired to make CBCT spectral imaging widely doable in practical application. 
\par Thus, this work aims to promote a source-detector joint multi-energy spectral imaging solution which takes advantages of both the FKS and DL-FPD, and to conduct a feasibility study on the first tabletop CBCT system with such a capability developed. Generally speaking, a joint multi-energy spectral solution can be enabled by using two or more types of spectral imaging technologies together\cite{tivnan2020combining,li2020mdm,chu2013combination,holbrook2020dual,yu2018dual,tao2020multi,wang2022fast}. In the literature, Tivnan $et$ $al$ conducted a comparative study of spectral imaging performance among standalone methods such as FKS, DL, and spatial spectral filter, and hybrid methods with different kinds of combinations, where simulation results showed that the hybrid acquisition strategies can achieve a better spectral diversity with a significant improvement in material decomposition performance as well, especially for enhanced contrast imaging task\cite{tivnan2020combining}. Sen $et$ $al$ optimized the X-ray source using the FKS technology in photon-counting detector based spectral imaging, where simulation results showed that the FKS can significantly suppress material decomposition noise when compared with the single kV scan. Additionally, noise can be further reduced by using the K-edge filter in the FKS scan\cite{wang2022fast}.
\par CBCT joint spectral imaging using the FKS and DL-FPD has not been reported with physics experiments to validate the feasibility. To close this gap, we develop the first tabletop multi-energy CBCT system by integrating DL-FPD and a FKS compatible high voltage generator, and conduct a set of physics experiments using the Gammex multi-energy phantom and the Kyoto head phantom. A Cramér-Rao Lower Bound (CRLB) based noise analysis for material decomposition is performed to better understand differences in the spectral imaging performance among the standalone CBCT spectral imaging technologies and the joint one. To compensate for the slightly angular mismatch between low- and high-energy projections in FKS, a dual-domain projection completion scheme is also implemented. 
\par The rest of this paper is organized as follows. Section \ref{sec:methods} presents the development of CBCT joint multi-energy tabletop system and the associated CRLB noise analysis. A set of physics phantom experiments are provided in sections \ref{sec:results}. Finally, we summarize this work in section \ref{sec:discussion} with a brief discussion.

\section{Materials and methods}
\label{sec:methods}
\subsection{Development of a joint spectral imaging system}

\par
\begin{figure}[h]
	\centering
	\includegraphics[width=\columnwidth]{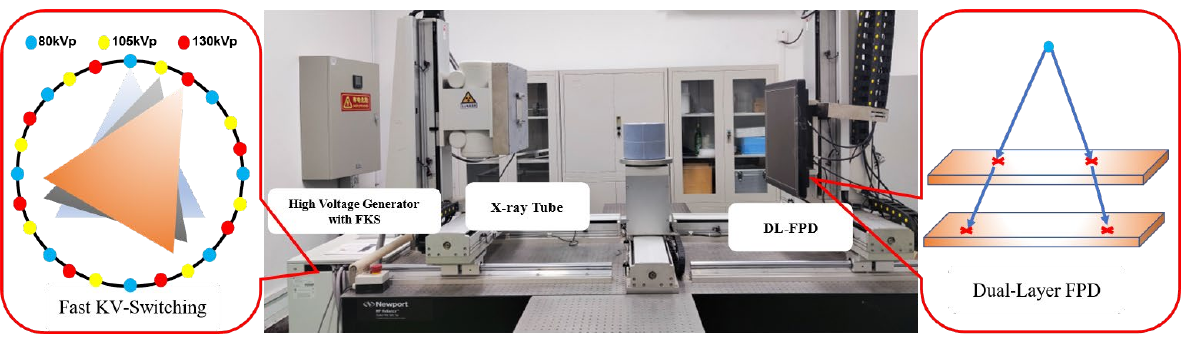}
	\caption{The developed tabletop CBCT system integrated the FKS and DL-FPD together. The high voltage generator is capable of FKS within the range of 40kV to 150kV at a switching frequency below 30Hz.}
	\label{fig:systemconfiguration}
\end{figure}

 As shown in Fig.\ref{fig:systemconfiguration}, the tabletop CBCT system with the joint spectral imaging capability was developed by integrating the FKS and DL-FPD together. In this work, the high voltage generator (EMD EPS50, Saint-Eustache, Quebec, Canada) was programmed to allow FKS according to the protocol provided by the manufacturer. This high voltage generator is capable of FKS within the range of 40kV to 150kV at a switching frequency below 30Hz, but with limitation that a single switching step cannot exceed +25kV. As a result, we operated the X-ray tube (Varex Imaging G-242, Salt Lake City, UT, USA) in a 80/105/130 kVp switching pairs. The DL-FPD (Varex Imaging XRD 4343RF, Salt Lake City, UT, USA) consists of two amorphous silicon (a-Si) panels with 2880 $\times$ 2880 pixels matrix corresponding to an illumination field of 43 $\times$ 43 $\mathrm{cm^2}$. Both the top and bottom panels were deposited with a $550$-$\mu m$-thick CsI scintillator with no intermediate metal filter in-between. To the best of our knowledge, this is the first tabletop CBCT system that integrates the FKS and DL-FPD. This joint multi-energy spectral imaging system can generate spectral projection data set with at least 4 different effective energy levels of both FKS and DL-FPD. Compared with the standalone CBCT spectral imaging approach, such a joint solution is able to make the spectral information of projection data more diversified, promising to improving energy separation for a better material decomposition. A detailed comparison will be provided later in section \ref{sec:results}.

\subsection{Spectral projection in the CBCT joint spectral imaging}

\par In the absence of scatter, the acquired spectral projection data can be expressed, 

\begin{equation}
	p^{j}_{i} = -\ln \left( N_{i_0}\int S_{i}(E)e^{-\sum_j \mu_f(E)T_{f}}R_j(E)e^{-\int_{\Omega}\mu(E,\mathbf{r})dl} dE\right)
	\label{eq:primary}
\end{equation}

\noindent where, ${N_{i_0}}$ represent the initial photon intensity emitted by the X-ray source; $S_{i}(E)$ is the normalized effective spectrum with $i$ indicating the low or high kV spectrum. $\mu_f(E)$ is the linear attenuation coefficient(LAC) of pre-detector filtrations. $T_f$ is the corresponding thickness. Here, $j$ indicates whether the projection data comes from the top- or bottom- layer of the DL-FPD; When $j = \rm{top}$, then $T_f = 0$; when $j = \rm{bottom}$, $T_f$ is the thickness of top-layer attenuation material; $R_j(E)$ is the detector response function; and $\mu(E,\mathbf{r})$ is the  of the scanned object at energy $E$ and location $\mathbf{r}$. 

\par
\begin{figure}[ht]
	\centering
	\includegraphics[width=11cm]{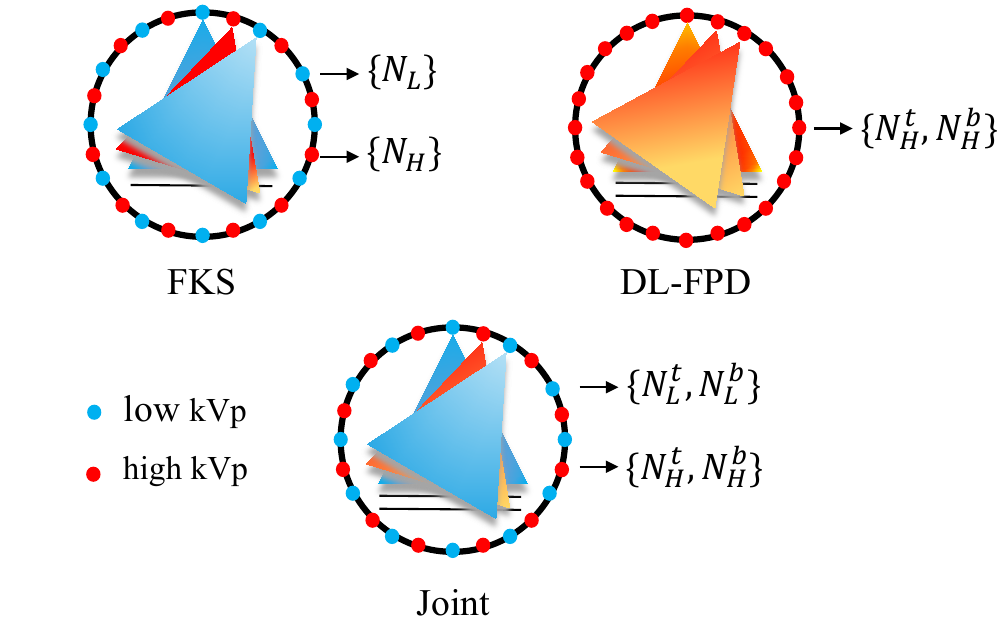}
	\caption{The schematic diagram of data acquisition for the FKS, DL-FPD and joint spectral imaging methods. For the FKS and joint spectral imaging method, a slightly angular mismatch between low- and high- energy projections.}
	\label{fig:projectiondistribution}
\end{figure}

 The joint spectral imaging solution utilizes a DL-FPD to acquire at least 4 energy levels of projection data between low- and high- kV switching, whose distribution characteristics compared with that of the FKS and DL-FPD are shown in Fig.\ref{fig:projectiondistribution}. For the sake of simplicity, we assume that the FKS alternates between high kV and low kV directly with no intermediate stage. The expected signal from the joint solution can be written as :

\begin{equation}
	N_{i}^{t} = {N_i}_0 \int S_{i}(E)R_t(E)e^{-\int_{\Omega}\mu(E,\mathbf{r})dl} dE
	\label{eq:Inten_top}
\end{equation}

\begin{equation}
	N_{i}^{b} = {N_i}_0 \int S_{i}(E)e^{-\sum_{j=b}\mu_f(E)T_f}R_b(E)e^{-\int_{\Omega}\mu(E,\mathbf{r})dl} dE
	\label{eq:Inten_bottom}
\end{equation}

\noindent where, $N_{i}^{t}$ and $N_{i}^{b}$ represent the low or high kV primary intensities after object attenuation which is obtained from the top layer and bottom layer of DL-FPD, respectively.

\par As shown in Fig.\ref{fig:projectiondistribution}, the projection sets of joint solution shows a slightly angular mismatch between low- and high- energy projections, which is similar to the projection of the regular FKS. The acquired data sets $\left\lbrace N_L^t,N_L^b,N_H^t,N_H^b \right\rbrace $ of joint solution contains more additional energy levels of projection data from the bottom layer of the DL-FPD compared with the data set $\left\lbrace N_L^t,N_H^t \right\rbrace$ from the regular FKS, which enhances the spectral information and photon detection efficiency in our developed tabletop CBCT system. While compared with the data set $\left\lbrace N_H^t,N_H^b \right\rbrace$ from the regular DL-FPD, the acquired projections for joint solution can be considered as substituting the part of high- energy projections of the DL-FPD with the dose-equivalent low- energy ones from the FKS, which makes the spectral information more diversified. 

\par For CT spectral imaging, it is assumed that the LAC can be decomposed as the sum of $M$ basis functions that are separable in energy- and space- domain as following, 
\begin{equation}
	\mu(E,\mathbf{r}) = \sum_{m=1}^{M} \alpha_{m}\left( \mathbf{r}\right) f_{m}\left( E\right) 
	\label{eq:LACModel}
\end{equation}
\noindent where $f_m$ are some known energy-dependent basis functions and $\alpha_{m}$ are the corresponding location-dependent coefficients. Two most commonly used basis functions are photoelectric absorption and Compton scattering, and basis material pair such as water and iodine or bone. For medical applications, the basis material method is typically used so that the LAC of scanned object can be simplified as 
\begin{equation}
	\mu(E,\mathbf{r}) = \alpha_{1}(\mathbf{r})\mu_{1}(E) + \alpha_{2}(\mathbf{r})\mu_{2}(E)
	\label{eq:basismaterialMu}
\end{equation}
\noindent with $\mu_1,\mu_2$ being the mass attenuation coefficients of basis material 1 and basis material 2. By substituting Eq.$\left( \ref{eq:basismaterialMu}\right)$ into Eq.$\left( \ref{eq:primary}\right) $, one gets,

\begin{equation}
	p^{j}_{i}  = -\ln\left(  N_{i_0} \int S_{i}(E)e^{-\sum_j \mu_f(E)T_f}R_j(E)e^{-A_1\mu_1(E)-A_2\mu_{2}(E)} dE\right) 
	\label{eq:rewriteprimary}
\end{equation}
\begin{equation}
	A_1 = \int_{\Omega}\alpha_{1}(\mathbf{r})dl,\; A_2 = \int_{\Omega}\alpha_{2}(\mathbf{r})dl
	\label{eq:passlength}
\end{equation}
\noindent with $A_1$ and $A_2$ being the corresponding material thickness (i.e., the line-integral of material density map). The purpose of material decomposition in projection domain is to inversely calculate $A_1 $ and $ A_2$ from the projection set $\left\lbrace p_i^j\right\rbrace $. That is, 
\begin{equation}
	\mathbf{A} = \mathbf{H}\left( \mathbf{p}\right)
	\label{eq:MDmap} 
\end{equation}
\par Here, $\mathbf{H}$ is the material decomposition matrix. The commonly used material decomposition methods in projection space include the polynomial function based direct mapping and the iterative ones such as the maximum-likelihood (ML) method.

\subsection{Benefits analysis for joint spectral imaging}

\par For joint spectral imaging, we can obtain 4 sets of spectral projection data $\left\lbrace N_L^t,N_L^b,N_H^t,N_H^b\right\rbrace $. Assuming $\left\lbrace N_L^t,N_L^b,N_H^t,N_H^b\right\rbrace $ follow the poisson distributions with the probability density functions being :
\begin{equation}
	p(N_i^j) = \frac{\overline{N_i^j}(\mathbf{A})^{N_i^j}}{N_i^j!}e^{-\overline{N_i^j}(\mathbf{A})}
	\label{eq:poissonPDF}
\end{equation}

\begin{equation}
	\overline{N_i^j}(\mathbf{A}) =  N_{i_0} \int S_i(E)e^{-\sum_j\mu_f(E)T_f} R_j(E) e^{-\mu_w(E)A_w-\mu_I(E)A_I} dE
	\label{eq:N_i}
\end{equation}

\noindent where, $\overline{N_i^j}(\mathbf{A})$ is the mean value of X-ray photons $N_i^j$. For the two estimated thickness ($A_w$ and $A_I$), their joint negative log-likelihood function can be given as :

\begin{eqnarray}
	&\mathbcal{L}(A_w,A_I|N_L^t,N_L^t,N_H^t,N_H^b) = -\ln \prod_{i,j} p_{i}^j \nonumber \\
	&= \sum_{i,j} \left[ \overline{N_i^j}(\mathbf{A})-N_{i}^j\ln(\overline{N_i^j}(\mathbf{A}))+\ln(N_{i}^j!)\right]   \label{eq:ML}
\end{eqnarray}

\par Here, the CRLB is calculated, which is widely used to estimate the achievable noise level of material decomposition as it can predict the minimum variance of an unbiased estimator such as the ML estimator.

\begin{equation}
	F_{\alpha\beta} = E\left[ -\frac{\partial^{2}\mathbcal{L}}{\partial{A_{\alpha}}\partial{A_\beta}}\right] 
	=\sum_{i,j}\frac{1}{\overline{N_i^j}(\mathbf{A})}\frac{\partial{\overline{N_i^j}(\mathbf{A}})}{\partial{A_\alpha}}\frac{\partial{\overline{N_i^j}(\mathbf{A}})}{\partial{A_\beta}}
	\label{eq:fisherinfo}
\end{equation}

\begin{equation}
	\frac{\partial{\overline{N_i^t}(\mathbf{A})}}{\partial{A_w}} =  N_{i_0} \int {A_wS_i(E)R_t(E)e^{-\mu_w(E)A_w-\mu_I(E)A_I}dE}
	\label{eq:partial_top}
\end{equation}

\begin{equation}
	\frac{\partial{\overline{N_i^b}(\mathbf{A})}}{\partial{A_w}} =  N_{i_0} \int {A_wS_i(E)e^{-\sum_j\mu_f(E)T_f}R_b(E)e^{-\mu_w(E)A_w-\mu_I(E)A_I} dE}
	\label{eq:partial_bottom}
\end{equation}

\begin{equation}
	\sigma^2_{A_i} \geq \mathrm{CRLB}_{A_i} = \frac{F_{i,i}}{|\mathbf{F}|}
	 \label{eq:CRLB_i}
\end{equation}

\par Here, $\sigma_{A_i}$ is the standard deviation of the estimated thickness of material $i$; $|\mathbf{F}|$ is the determinant of the Fisher information matrix $\mathbf{F}$ with $F_{ij}$ being its elements. For the same material decomposition task, using a standalone spectral imaging technology of either FKS or DL-FPD, the expression of CRLB will be similar, for which a detailed derivation can be found in reference\cite{roessl2009cramer}. 

\subsection{Spectral completion and reconstruction}
\par To achieve high quality material decomposition from multi-energy spectral projection data, the ML method was employed in this study. Theoretically, ML algorithm can achieve CRLB when the projection sets are noise independent. According to Eq.$\left( \ref{eq:ML}\right) $, the objective function for material decomposition can be expressed as,

\begin{equation}
	\mathbf{A}_{\rm{ML}} = arg\: {\displaystyle\min_{\mathbf{A}}} \sum_{i,j} \overline{N_i^j}(\mathbf{A})-N_{i}^j \cdot \ln(\overline{N_i^j}(\mathbf{A}))
	\label{eq:MLOBJ}
\end{equation}

\noindent where, $\mathbf{A}$ is the basis material thickness matrix. $N_i^j$ is the measured photons. For joint solution, $N_i^j$ denotes one of the projection sets $\left\lbrace N_L^t,N_L^b,N_H^t,N_H^b\right\rbrace $. \par
\begin{figure}[h]
	\centering
	\includegraphics[width=10cm]{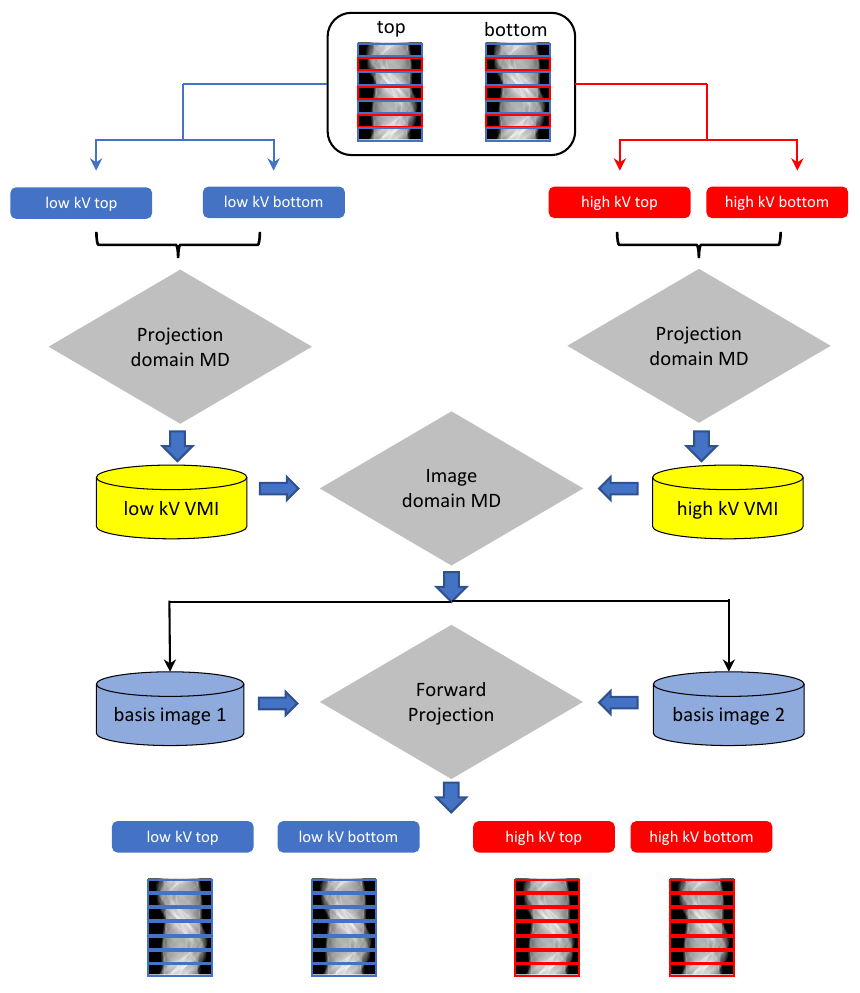}
	\caption{The workflow of dual-domain projection completion scheme. First, the projection of each kVp from the top- and bottom- layer is decomposed to obtain the optimal VMIs in the projection domain. Then the image-domain material decomposition is applied on the VMIs to obtain the basis material images. Finally, a forward projection is implemented to compensate high- and low- energy projection.}
	\label{fig:dualdomain}
\end{figure}

 However, there is only one kVp projection data available for a specific voltage for a given view due to the use of FKS. Similar to the regular FKS spectral imaging, a projection view interpolation is necessary if a projection-domain material decomposition is desired. In this study, we proposed a dual-domain (i.e., both image and projection domains) projection completion  (view interpolation) algorithm to compensate for the slightly angular mismatch between high- and low- energy projections as shown in Fig.\ref{fig:dualdomain}. 

\par This algorithm is primarily divided into two parts. Firstly , the projection-domain material decomposition $\boldsymbol{M^{-1}_P}$ and reconstruction operator $\boldsymbol{R}$ are applied to obtain the VMIs for low-kVp negative-logarithm projection $\left\lbrace  \mathnormal{p_L^t,p_L^b} \right\rbrace $ and high-kVp negative-logarithm projection $\left\lbrace  \mathnormal{p_H^t,p_H^b} \right\rbrace$, respectively. 
\begin{equation}
	\rm{VMI}_L = \boldsymbol{R} \left( \boldsymbol{M^{-1}_P} \left( \mathnormal{ p_L^t,p_L^b}\right) \right) , \quad \rm{VMI}_H = \boldsymbol{R} \left( \boldsymbol{M^{-1}_P} \left( \mathnormal{p_H^t,p_H^b}\right) \right)
	\label{eq:MD_1}
\end{equation}

 Then the image-domain material decomposition $\boldsymbol{M^{-1}_I}$ is performed on the VMIs $\left\lbrace \rm{VMI}_L,\rm{VMI}_H \right\rbrace $ to generate the basis material images $\left\lbrace \rm{B_1,B_2} \right\rbrace $ free from beam hardening artifacts.

 \begin{equation}
 	\rm{B_1,B_2} = \boldsymbol{R} \left( \boldsymbol{M^{-1}_I}\left( 	\rm{VMI}_L,	\rm{VMI}_H \right) \right)
 	\label{eq:MD_2}
 \end{equation}
 Subsequently, a forward projection $ \mathbcal{F}$ is implemented to estimate high- and low- energy projection. Finally, the completed projections $N_{\rm{completed}}$ are used for a final material decomposition with the ML algorithm, from which both the basis material images and VMIs can be reconstructed using a conventional reconstruction such as the FDK\cite{feldkamp1984practical}.
 \begin{equation}
 	N_{\rm{completed}} = \mathbcal{F}\left(\rm{
 		B_1,B_2} \right)
 	\label{eq:MD_2}
 \end{equation}

\subsection{Experimental study}
\par To quantitatively compare the performance of joint solution with individual solution via FKS or DL-FPD, a numerical simulation experiment was conducted. Water and iodine were selected as the basis material set and their CRLBs were evaluated. A DL-FPD with CsI scintillator crystal was used, with both layers having a thickness of $ T_f = 550\mu m$ and no intermediate filter. 80/140 kVp pair was used in the numerical simulation. To make a fair noise performance comparison, a projection-based dose assessment method was used in the simulation to keep dose balanced between low and high kV scans\cite{tian2013projection}.

\par We validated the joint spectral imaging performance on our developed tabletop CBCT as shown in Fig.\ref{fig:systemconfiguration}. The high voltage generator provides consecutive x-ray pulses that alternate different kVps using a programmed sequence. However, due to the limitation that the single switch cannot exceed 25kV, we alternated kV among three voltages (80/105/130 kVps) to achieve sufficient energy separation in our study. For simplicity, so far only the projection data from 80/130 kVp scans were used in this study. Similar to the dose assessment index used in simulation, we employed a projection-based dose metric to assess the effective dose of low and high kVp scan configurations\cite{tian2013projection}. Dose balance is ensured by adopting different mAs parameters between low and high kVp scan configurations. The Gammex multi-energy phantom (Sun Nuclear Corporation, Middleton, WI) with different concentrations rods of $2,5,10,15 \, \rm{mg/cc}$ and the head phantom (Kyoto Kagaku, Kyoto, Japan) are scanned to achieve a quantitative and realistic comparative study. Both the cone-beam and fan-beam scans were conducted. For fan-beam experiments, the X-ray beam is narrow-collimated to eliminate scatter. For cone-beam experiments, scatter correction has to be done carefully. Here, we use the CBCT Software Tools (CST 2.2, Varex Imaging) which estimates the scatter distribution by solving the linear Boltzmann transport equation\cite{maslowski2018acuros}. The cone-beam artifact reduction (CBAR) and the low signal correction method is applied in the head experiments\cite{forthmann2009adaptive,wang2023dual}. The spectral imaging experimental parameters are summarized in Table.\ref{table:geo_param}.

\subsection{Evaluations and metrics}

\par Energy separation is a good indicator to roughly assess spectral imaging performance from a CT scan, which is defined as the difference between the effective energies of two spectra,
\begin{equation}
	\Delta E = \frac{\int E\cdot S_1(E) dE}{\int S_1(E) dE} - \frac{\int E\cdot S_2(E)dE}{\int S_2(E)dE}
	\label{eq:energyseparation}
\end{equation}

\par To characterize the quantitative performance of basis image after material decomposition, the contrast-to-noise ratio (CNR) in the selected regions-of-interest (ROI) is also calculated as 
\begin{equation}
	\mathrm{CNR} = \frac{ \left| \mu_i - \mu_{ref}\right| }{\sqrt{\frac{1}{2}\left( \sigma_i^2+\sigma_{ref}^2\right)}}
	\label{eq:CNR}
\end{equation}
\noindent where, $\mu$ is the mean value of the ROI; and $\sigma$ is the standard deviation. 

\par The nonuniformity of the reconstructed CT image is measured as the maximum difference of among the selected ROIs within supposedly uniform regions,

\begin{equation}
	\Delta = \rm{max}\left(\mu_i\right) - \rm{min}\left(\mu_i\right)
	\label{eq:NU}
\end{equation}

\par The mean-relative-error (MRE) is used to evaluate the accuracy bias of the total ROIs , which is defined as, 
\begin{equation}
	\mathrm{MRE} =\frac{1}{N} \cdot \sum_{i}^{N}  \frac{\left| \mu_i-\mu_{ref} \right| }{\mu_{ref}} \times 100\%
	\label{eq:MRD}
\end{equation}

\begin{table*}
	\centering \caption{The tabletop CBCT experiment Parameters.}
	\label{table:geo_param}
	\begin{tabular}{cc}
		\toprule
		Parameter                               &Value\\
		\midrule
		Source-isocenter distance               &750mm	\\
		Source-detector distance              	&1184mm  \\
		Detector size                           & $0.45 \times0.45 \rm{mm^2}$\\
		&960$\times$960 pixel matrix\\
		Scintillator                            &CsI\\
		Scintillator thickness of top layer		&550 $\rm{\mu m}$	\\
		Scintillator thickness of bottom layer	&550 $\rm{\mu m}$	\\
		Intermediate filter						& None\\
		Source filter 							&2mm Al + 0.4mm Cu \\
		kVp/mAs									&80kVp/500mAs\\
		&105kVp/400mAs \\
		&130kVp/160mAs \\
		kV-switching speed 						&20fps \\
		Acquisition frames						&1500 frames/360\degree\, for FKS and Joint\\
												&1000 frames/360\degree\, for DL-FPD \\
		Phantom									&multi-energy phantom \& head phantom	\\
		Recon kernel							&Identity for multi-energy phantom	\\
												&Smooth for head phantom	\\ 
		Recon size								&$512 \times 512$ for fan-beam\\
												&$512 \times 512 \times 512$ for cone-beam\\
		Voxel size								&$0.5 \times 0.5 \rm{mm^2}$ for fan-beam \\
												&$0.5 \times 0.5 \times 0.5 \rm{mm^3}$ for cone-beam	\\
		\bottomrule
	\end{tabular}
\end{table*}

\section{Results} 
\label{sec:results}
\subsection{CRLB calculation}
\begin{figure}
	\centering
	\includegraphics[width=12cm]{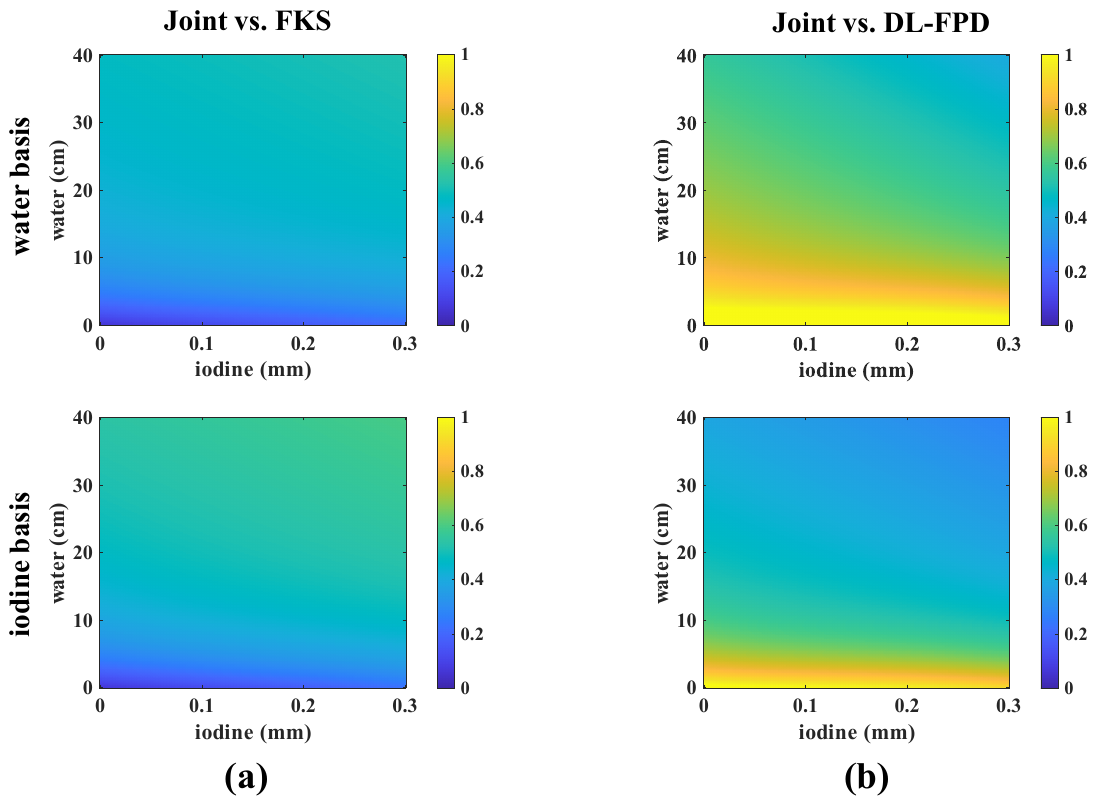}
	\caption{The CRLB ratio between joint solution and the FKS or DL-FPD, the ratio less than 1 indicates that the joint solution has lower CRLB. The first row is for water while the second row for iodine. Compared with the FKS, the joint solution has the lower CRLB throughout the thickness combinations in the simulation. Compared with the DL-FPD, the joint solution has the lower CRLB in most cases. When the water thickness is small, the CRLB of the joint solution and the DL-FPD is comparable.}
	\label{fig:CRLB}
\end{figure}

\begin{figure*}[ht]
	\centering
	\includegraphics[width=14cm]{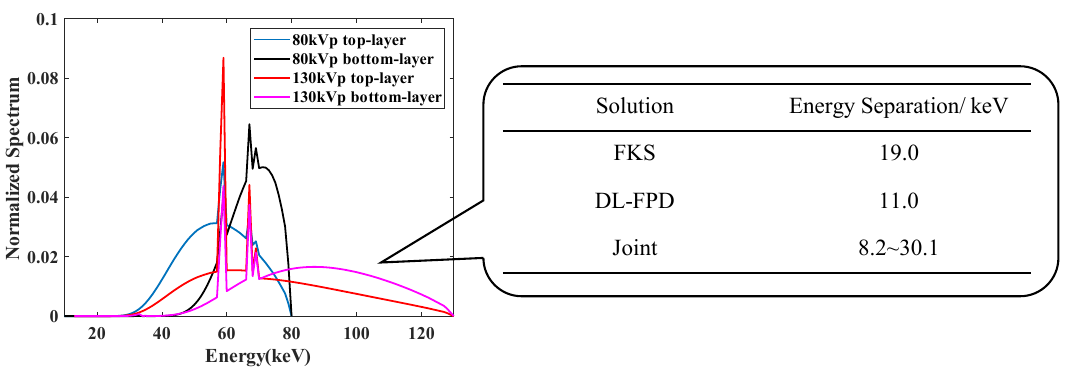}
	\caption{The estimated effective spectra for 80/130 kVp pair and different layers in tabletop CBCT system. The energy separation after 20-cm water attenuation is calculated for different spectral imaging solutions. The energy separation of joint spectral imaging is calculated by the 4 spectra. The smallest one 8.2 keV is calculated from the 80kVp top-layer spectrum and 80kVp bottom-layer spectrum. The largest one 30.1 keV is calculated from the 80kVp top-layer spectrum and 130kVp bottom-layer spectrum. }
	\label{fig:TabletopSpectrumAndEnergySeparation}
\end{figure*}

\par Figure.\ref{fig:CRLB} illustrates the CRLB ratio for the joint solution versus the FKS and the DL-FPD, respectively. One can observe that the joint solution has a lower noise bound compared with the FKS, which is due to the fact that the bottom layer of the DL-FPD increases the detection efficiency and improves energy separation at the same time. While compared with the DL-FPD as shown in Fig.\ref{fig:CRLB}(b), joint solution achieves lower noise bounds for most combinations of water and iodine thickness sets. However, there are few specific combinations of basis material sets where the CRLB of the joint solution could be higher than that of the DL-FPD. These situations mainly arise when the thickness of water is relatively small (less than 3 cm in our simulation), meaning that the advantages of the joint solution over the DL-FPD are not apparent when the scanned object is very small. 

\subsection{Physics experiment on the developed tabletop CBCT system}

\par The normalized effective spectra are estimated by a series of wedge experiments with filters using different materials and thicknesses\cite{sidky2005robust}. Fig.\ref{fig:TabletopSpectrumAndEnergySeparation} shows the estimated spectra and the calculated energy separation for different spectral imaging solutions using Eq.$\left( \ref{eq:energyseparation}\right) $. Compared with the FKS and DL-FPD, the joint solution has the maximum energy separation of 30.1 keV, which is achieved by the 80kVp top-layer spectrum and the 130kVp bottom-layer spectrum. A projection-based dose assessment method is adopted to keep dose balanced between the low kV scan and high kVp scan\cite{tian2013projection}.

\subsubsection{Multi-energy phantom experiments\rm{ :}}
 \par
\begin{figure*}[h]
 	\centering
 	\includegraphics[width=10cm]{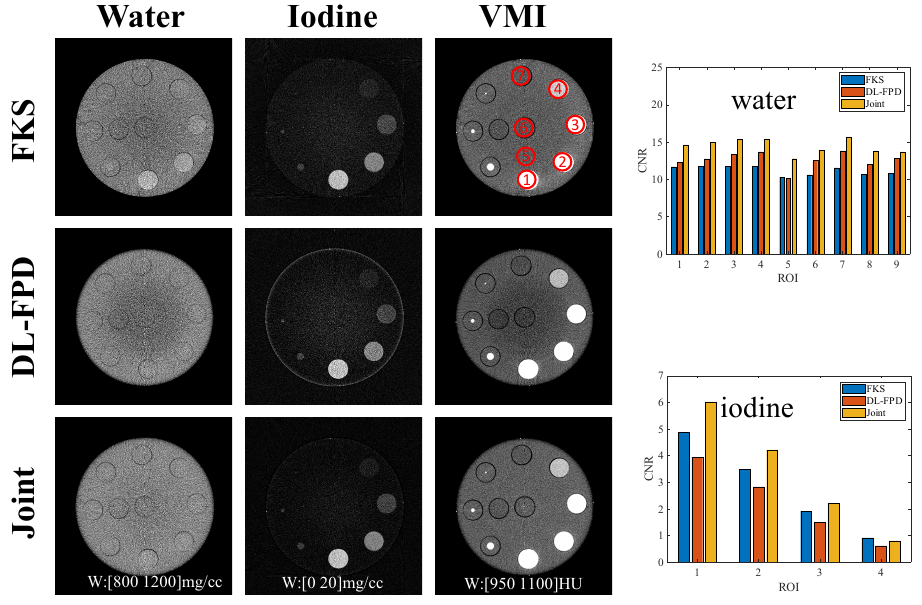}
 	\caption{The multi-energy phantom material decomposition reconstruction results for fan-beam experiment. The VMIs at 70keV are displayed here. The ROIs are selected as shown in VMI. The CNR figures of water and iodine images of multi-energy phantom in fan-beam experiment are shown in the bottom row.  }
 	\label{fig:MEPhanFB}	
 \end{figure*}

\begin{figure}[h]
 	\centering
 	\includegraphics[width=10cm]{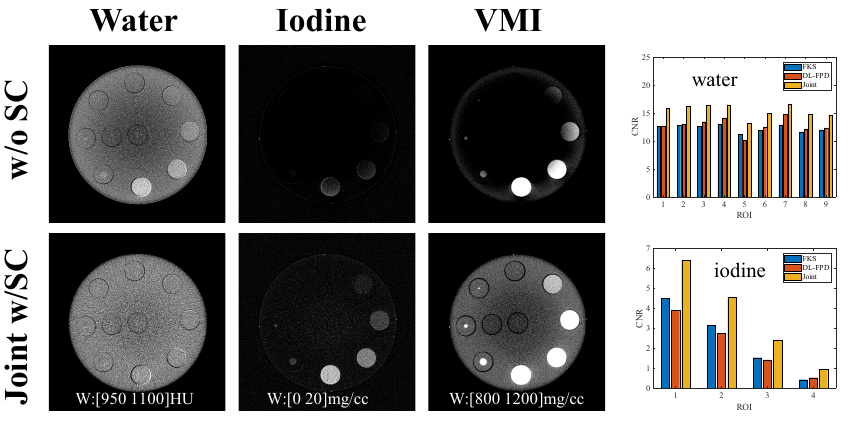}
 	\caption{The material decomposition result of multi-energy phantom for cone-beam experiment. The VIMs at 70keV are displayed here. W (Wo)/SC : with (without) scatter correction. The CNR of water and iodine images of multi-energy phantom for cone-beam experiment are shown in right side.}
 	\label{fig:MEPhanCB}
 \end{figure}

\begin{figure}[h]
 	\centering
 	\includegraphics[width=8cm]{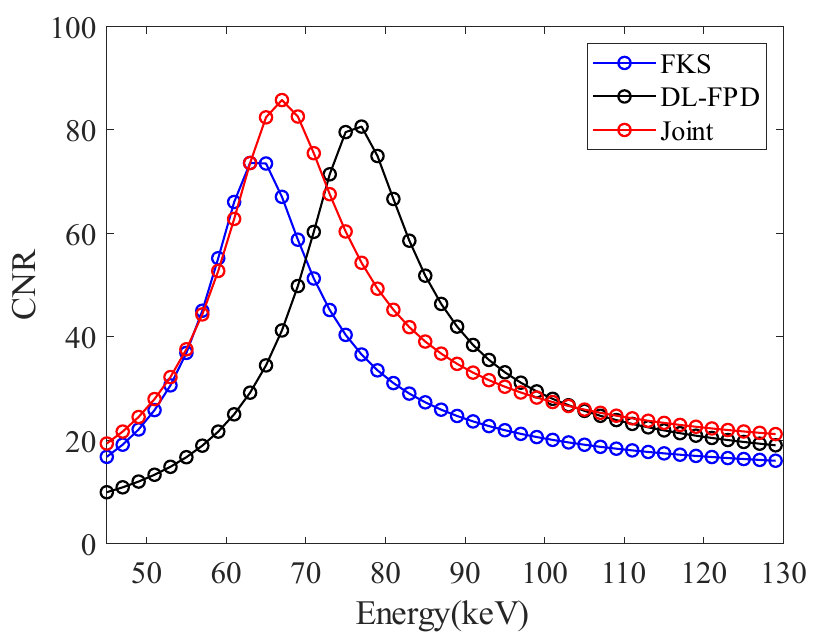}
 	\caption{The CNR for 10 mg/cc iodine of multi-energy phantom as a function of VMI energy in cone-beam experiment. The energies at the optimal CNR are 63keV for FKS, 77keV for DL-FPD and 67keV for joint spectral imaging, respectively.}
 	\label{fig:CNRIodine20}
 \end{figure}

 For the multi-energy phantom study, the reconstructed basis material images and VMIs for fan-beam and cone-beam experiments are shown in Fig.\ref{fig:MEPhanFB} and Fig.\ref{fig:MEPhanCB}, respectively. The reference and measured concentrations for basis material reconstructed image are summarized in Table.\ref{table:waterFBHU} and Table.\ref{table:waterCBHU}. In the reconstructed images, the MRE of water concentration are 114.8, 80.49, 51.98 in fan-beam study and 48.9, 35.87, 19.98 in cone-beam study for the FKS, DL-FPD and joint spectral imaging respectively; The MRE of iodine concentration are 0.74, 0.26, 0.26 in fan-beam study and 1.82, 0.18, 0.31 in cone-beam study for FKS, DL-FPD and joint spectral imaging respectively. The CNR of basis material images of multi-energy phantom for fan-beam study and cone-beam study are shown in the right side of Fig.\ref{fig:MEPhanFB} and Fig.\ref{fig:MEPhanCB}, respectively. The measurements were made for each inserted iodine contrast rods. The background values are obtained from a circular region outside the phantom. Compared with the FKS and the DL-FPD, the CNR of water image for joint spectral imaging are boosted by an average improvement of $29.2\%$, $15.2\%$ in fan-beam study and $14.5\%$, $25.6\%$ in cone-beam study respectively; The CNR of iodine image are boosted by an average improvement of $11.8\%$, $45.1\%$ in fan-beam study and $53.8\%$, $80.5\%$ in cone-beam study respectively. The CNR for 10 $\rm{mg/cc}$ iodine rod as a function of VMI energy in cone-beam study is shown in Fig.\ref{fig:CNRIodine20}. The best CNR is 73.5 at 65keV for FKS, 80.6 at 77keV for DL-FPD and 85.8 at 67keV for joint spectral imaging, respectively.
 
 \begin{table*}
 	\centering
 	\caption{Material decomposition accuracy comparison of the selected ROIs in Fig.\ref{fig:MEPhanFB} among FKS, DL-FPD and joint spectral imaging solutions for fan-beam experiment.}
 	\label{table:waterFBHU}
 	\begin{tabular}{lllllllllll}
 		\toprule
 		\multicolumn{2}{c}{ROI} & 1 & 2 & 3 & 4 & 5 & 6 & 7  & MRE \\
 		\midrule
 		\multirow{4}{*}{water} &Reference(mg/cc)  &&& 1000 \\
 		\cmidrule{2-10}
 		&FKS &1075 &1050 &1025 &1006 &957.6 &974.8 &990.1  & 3.34\%  \\
 		&DL-FPD  &1020 &1017 &1015 &1008 &957.4 &969.4 &1001 	&1.95\% \\
 		&Joint &1034 &1028 &1019 &1017 &992.5 &1004 &1008 &1.68\%  \\
 		\midrule
 		\multirow{4}{*}{iodine}&Reference(mg/cc) &15 &10 &5 &2	\\
 		\cmidrule{2-10}
 		&FKS &14.8 &10.2 &5.44 &2.52  &\quad\textbackslash &\quad\textbackslash &\quad\textbackslash  &9.53\%\\
 		&DL-FPD  &15.2 &10.2 &4.97 &1.96	&\quad\textbackslash &\quad\textbackslash &\quad\textbackslash  &1.48\%\\
 		&Joint &14.9 &9.94 &4.96 &1.78 &\quad\textbackslash &\quad\textbackslash &\quad\textbackslash  &3.26\%\\
 		\bottomrule
 	\end{tabular}
 \end{table*}
 
 \begin{table*}[htp]
 	\centering
 	\caption{Material decomposition accuracy comparison of the multi-energy phantom in Fig.\ref{fig:MEPhanFB} among FKS, DL-FPD and joint spectral imaging solutions for cone-beam experiment.}
 	\label{table:waterCBHU}
 	\begin{tabular}{lllllllllll}
 		\toprule
 		\multicolumn{2}{c}{ROI} & 1 & 2 & 3 & 4 & 5 & 6 & 7 & MRE \\
 		\midrule
 		\multirow{4}{*}{water} &Reference(mg/cc)  &&&& 1000 \\
 		\cmidrule{2-10}
 		&FKS &1095 &1074 &1050 &1039 &1013 &1029 &1028 & 4.69\%  \\
 		&DL-FPD  &1026 &1023 &1019 &1015 &935.7 &959.1 &1004 &2.75\% \\
 		&Joint &1035 &1025 &1014 &1009 &970.6 &990.1 &1006 &1.83\%  \\
 		\midrule
 		\multirow{4}{*}{iodine} &Reference(mg/cc) &15 &10 &5 &2	\\
 		\cmidrule{2-10}
 		&FKS &12.26 &8.14 &3.87 &1.01 &\quad\textbackslash &\quad\textbackslash &\quad\textbackslash &27.24\%\\
 		&DL-FPD &14.94  &10.04 &4.84 &1.69 &\quad\textbackslash &\quad\textbackslash &\quad\textbackslash &4.87\% \\
 		&Joint &14.41 &9.82 &5.04 &1.95 &\quad\textbackslash &\quad\textbackslash &\quad\textbackslash &2.26\% \\
 		\bottomrule
 	\end{tabular}
 \end{table*}

\begin{figure*}[h]
	\centering
	\includegraphics[width=16cm]{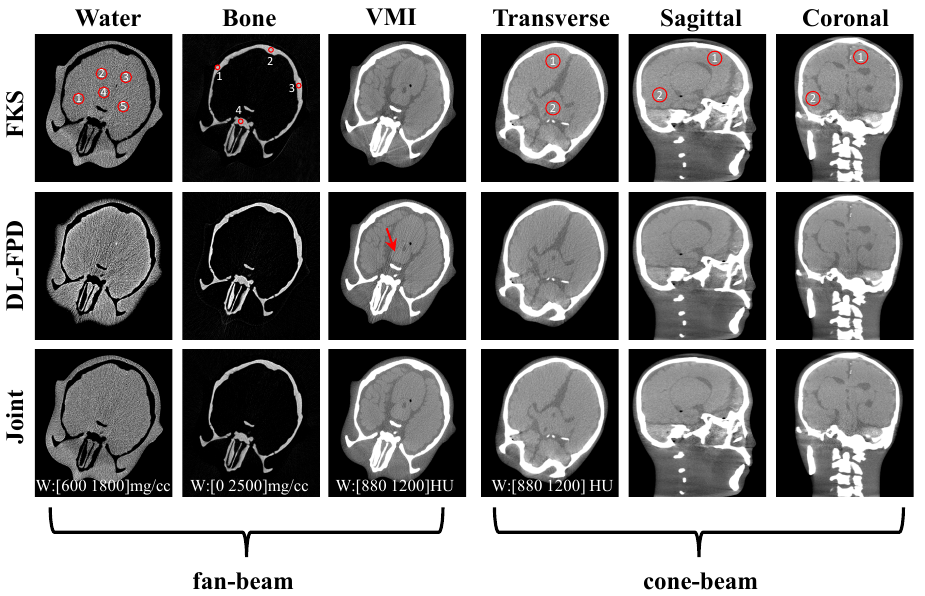}
	\caption{The head phantom material decomposition reconstruction results. The left side are the basis material images and VMIs for fan-beam experiment. And the right side are the 3D reconstruction results of VMIs for cone-beam experiment. The VMIs at 70keV are displayed here.}
	\label{fig:HeadFB}	
\end{figure*}

\subsubsection{Head phantom experiments\rm{ :}}
\par
\begin{figure}
	\centering
	\includegraphics[width=15cm]{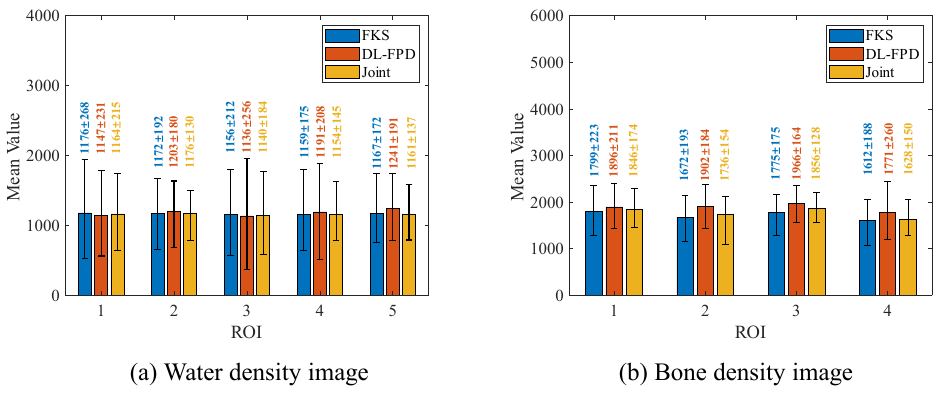}
	\caption{The measured mean value and standard deviation of the selected ROIs in Fig.\ref{fig:HeadFB} for fan-beam experiment. The joint spectral imaging has the lower noise in water and bone density basis images compared with the the standalone ones including the FKS and the DL-FPD technologies. Unit : $\rm{mg/cc}$.}
	\label{fig:HeadFBHU}
\end{figure}

 For the head phantom study, the material decomposition result for fan-beam and cone-beam experiments are shown in Fig.\ref{fig:HeadFB}. The measured mean value of the water and bone basis image for fan-beam experiment is shown in Fig.\ref{fig:HeadFBHU}. The joint spectral imaging has the lower noise in water and bone basis images compared with the the standalone ones including the FKS and the DL-FPD technologies. It is evident that the DL-FPD spectral imaging exhibits streaking artifacts in VMI whereas there is a significant improvement in the joint spectral imaging as the red arrow shown in Fig.\ref{fig:HeadFB}. This improvement is attributed to the strong x-ray attenuation and insufficient energy separation. Table.\ref{table:Measured3D} quantitatively presents the material decomposition result of the 3D reconstructed VMIs. Compared with the FKS and the DL-FPD, the standard deviation of the joint spectral imaging are reduced by an average decrement of $19.5\%$ and $8.1\%$ in cone-beam study, respectively. The CT number difference between the two ROIs selected in each plane is also reduced from 9.8, 18.3 to 7.3 HU for transverse plane, from 2.2, 9.9 to 1.2 HU for sagittal plane and from 3.4, 9.5 to 3.1 HU for coronal plane, respectively.

\begin{table*}[h]
	\centering
	\caption{The measured mean values and standard deviation of the selected ROIs in Fig.\ref{fig:HeadFB} for cone-beam experiment.}
	\label{table:Measured3D}
	\begin{tabular}{llccccc}
		\toprule
		\multirow{2}{*}{Plane} &\multirow{2}{*}{Solution} & \multicolumn{2}{c}{ROI 1} &\multicolumn{2}{c}{ROI 2} &\multirow{2}{*}{$\Delta$(HU)} \\
		&& Mean(HU) &Std(HU)	& Mean(HU) &Std(HU)	\\
		\midrule
		\multirow{3}{*}{Transverse} &FKS &1067.2 &9.8 &1059.6	&11.1 & 9.8 \\
									&DL-FPD &1062.7 &8.6 &1044.4	&10.9 & 18.3 \\
									&Joint &1067.1 &8.0 &1059.8	&9.3 & 7.3 \\
		\midrule
		\multirow{3}{*}{Sagittal} &FKS &1065.4 &8.0 &1067.6	&10.1 &2.2 \\
										   &DL-FPD &1069.3 &6.8 &1059.4	&8.5&9.9 \\
									       &Joint &1067.7 &6.7 &1066.5	&8.0 &1.2 \\
		\midrule
		\multirow{3}{*}{Coronal}  &FKS &1067.5 &7.3 &1070.9	&11.4 &3.4 \\
								  &DL-FPD &1071.0 &6.3 &1061.5	&9.7&9.5 \\
								  &Joint &1067.5 &5.9 &1070.6	&8.4 &3.1 \\
		\bottomrule	
	\end{tabular}
	
\end{table*}

\section{Discussion and conclusion}
\label{sec:discussion}

\par In this study, we validated the superiority of the joint spectral imaging method using the FKS and DL-FPD together through the theoretical analysis ands physics experiments. We derived the theoretical noise lower bound for joint spectral imaging using the CRLB calculation and demonstrated that the joint solution outperforms the standalone ones including the FKS and the DL-FPD technologies, in terms of noise performance for most cases in simulation. Subsequently, we conducted a set of physics experiments on a developed tabletop CBCT, where the FKS and the DL-FPD technologies were first developed together, using both a multi-energy phantom and a head phantom. Quantitatively, the CNRs of joint solution showed an averaged improvement of 21.9$\%$, 20.4$\%$ for water image and 32.8$\%$ ,62.8$\%$ for iodine image compared with the FKS and DL-FPD technologies, respectively. From the head phantom experiment, it is evident that joint solution significantly reduces streaking artifacts compared with the standalone technologies.

\par As shown in Fig.\ref{fig:CRLB},  the noise performance of the joint solution is significantly superior to the DL-FPD solution when the thickness of water is relatively large. While for smaller water thickness, their noise performance are comparable. One possible reason for this phenomenon is that the energy separation of the FKS increases after being attenuated by scanned object and this would be more pronounced with the increase in object size, while for the DL-FPD, the opposite trend is observed. When the object is relatively small, the energy separation of the FKS does not increase significantly compared with the DL-FPD\cite{jin2014cnr}. However, low-energy X-ray photons attenuate more than high-energy X-ray photons, leading to increased statistical noise. The noise of material decomposition for the joint solution is comparable to, or slightly worse than that for the DL-FPD technology when the object size is very small. 

\par In the head phantom study, we observed that the joint spectral imaging method significantly reduced streaking artifacts. In single-energy CT imaging, photon starvation is a major cause of streaking artifacts. As indicated in Fig.\ref{fig:HeadFB}, joint solution exhibited a noticeable reduction in streaking artifacts compared with the DL-FPD technology. However, streaking artifacts in this work is not caused by the photon starvation. Because the joint spectral imaging can be viewed as replacing half of the high-energy projections in the DL-FPD technology with low-energy projections, which faces a more pronounced photon starvation effect due to stronger attenuation. Nevertheless, the joint solution reduces streaking artifacts indicating that streaking artifacts in spectral imaging are also influenced by energy separation. When energy separation is insufficient, material decomposition becomes unstable and more susceptible to streaking artifacts. This indirectly demonstrates the robustness and superiority of the joint spectral imaging.  

\par The joint spectral imaging solution also has limitations. As previously mentioned, its advantages are not prominent when the scanned object is small compared with the DL-FPD. Due to the presence of low-energy spectra, it may face issues like X-rays being unable to penetrate when dealing with large objects, which are less problematic with DL-FPD technology.

\par Due to the hardware limitations, we only realized a three-kVp switching pattern (low, high and intermediate step) and picked the low and high projection data with the projection data in our validation for simplicity. In the future, the intermediate step during the fast kV-switching can be avoided as the high voltage technology advanced. Of course, the multi-kV-switching pattern may also have benefits in some specific applications.
\par The joint spectral imaging method holds significant potential for applications in multi-material decomposition using a more flexible kVp-switching pairs. In this work, we only attempted the dual-material decomposition. In the future, we are very much interested in exploring the possibilities of multi-material decomposition for spectral imaging as well.

\section*{ACKNOWLEDGMENTS}
\par This project was supported in part by the National Key R$\&$D Program of China (No.2022YFE0131100), and the National Natural Science Foundation of China (No.12075130 and No.U20A20169).

\newcommand{\newblock}{}
\bibliographystyle{MedPhys}
\bibliography{DLref}

\begin{thebibliography}{10}
\newcommand{\enquote}[1]{``#1''}

\bibitem{mccollough2020principles}
C.~H. McCollough, K.~Boedeker, D.~Cody, X.~Duan, T.~Flohr, S.~S. Halliburton,
  J.~Hsieh, R.~R. Layman, and N.~J. Pelc, \enquote{Principles and applications
  of multienergy {CT}: Report of aapm task group 291,} Medical Physics
  \textbf{47}, e881--e912 (2020).

\bibitem{glazebrook2011identification}
K.~N. Glazebrook, L.~S. Guimar{\~a}es, N.~S. Murthy, D.~F. Black, T.~Bongartz,
  N.~J.~Manek, S.~Leng, J.~G. Fletcher, and C.~H. McCollough,
  \enquote{Identification of intraarticular and periarticular uric acid
  crystals with dual-energy {CT}: initial evaluation,} Radiology
  \textbf{261}(2), 516--524 (2011).

\bibitem{bongartz2015dual}
T.~Bongartz, K.~N. Glazebrook, S.~J. Kavros, N.~S. Murthy, S.~P. Merry, W.~B.
  Franz, C.~J. Michet, B.~M.~A. Veetil, J.~M. Davis, T.~G. Mason \emph{et~al.},
  \enquote{Dual-energy {CT} for the diagnosis of gout: an accuracy and
  diagnostic yield study,} Annals of the rheumatic diseases \textbf{74}(6),
  1072--1077 (2015).

\bibitem{mangesius2021dual}
S.~Mangesius, T.~Janjic, R.~Steiger, L.~Haider, R.~Rehwald, M.~Knoflach,
  G.~Widmann, E.~Gizewski, and A.~Grams, \enquote{Dual-energy computed
  tomography in acute ischemic stroke: state-of-the-art,} European Radiology
  \textbf{31}, 4138--4147 (2021).

\bibitem{van2021virtual}
F.~van Ommen, J.~W. Dankbaar, G.~Zhu, D.~N. Wolman, J.~J. Heit, F.~Kauw,
  E.~Bennink, H.~W. de~Jong, and M.~Wintermark, \enquote{Virtual monochromatic
  dual-energy {CT} reconstructions improve detection of cerebral infarct in
  patients with suspicion of stroke,} Neuroradiology \textbf{63}, 41--49
  (2021).

\bibitem{sun2022pulmonary}
W.~Sun, H.~Tan, Y.~Wang, A.~Xie, X.~Tan, P.~Liu, D.~Xu, and F.~Huang,
  \enquote{Pulmonary {CT} scans of white rabbits using the selective photon
  shield technique of the third-generation dual-source {CT},} Journal of
  Radiological Protection \textbf{42}(2), 021527 (2022).

\bibitem{harvey2019impacts}
E.~C. Harvey, M.~Feng, X.~Ji, R.~Zhang, Y.~Li, G.-H. Chen, and K.~Li,
  \enquote{Impacts of photon counting {CT} to maximum intensity projection
  ({MIP}) images of cerebral {CT} angiography: theoretical and experimental
  studies,} Physics in Medicine \& Biology \textbf{64}(18), 185015 (2019).

\bibitem{deng2022multi}
Y.~Deng and H.~Gao, \enquote{Design of scatter-decoupled material decomposition
  for multi-energy blended {CBCT} using spectral modulator with flying focal
  spot,} in: \emph{The 17th Virtual International Meeting on Fully 3D Image
  Reconstruction in Radiology and Nuclear Medicine}, 78--81 (2023).

\bibitem{tivnan2022design}
M.~Tivnan, W.~Wang, G.~Gang, and J.~W. Stayman, \enquote{Design optimization of
  spatial-spectral filters for cone-beam {CT} material decomposition,} IEEE
  Transactions on Medical Imaging \textbf{41}(9), 2399--2413 (2022).

\bibitem{cai2023benefits}
E.~Y. Cai, C.~De~Caro, K.~Treb, and K.~Li, \enquote{Benefits of using removable
  filters in dual-layer flat panel detectors,} Physics in Medicine \& Biology
  \textbf{68}(8), 085013 (2023).

\bibitem{gang2012cascaded}
G.~J. Gang, W.~Zbijewski, J.~Webster~Stayman, and J.~H. Siewerdsen,
  \enquote{Cascaded systems analysis of noise and detectability in dual-energy
  cone-beam {CT},} Medical Physics \textbf{39}(8), 5145--5156 (2012).

\bibitem{cassetta2020fast}
R.~Cassetta, M.~Lehmann, M.~Haytmyradov, R.~Patel, A.~Wang, L.~Cortesi,
  D.~Morf, D.~Seghers, M.~Surucu, H.~Mostafavi \emph{et~al.},
  \enquote{Fast-switching dual energy cone beam computed tomography using the
  on-board imager of a commercial linear accelerator,} Physics in Medicine
  Biology \textbf{65}(1), 015013 (2020).

\bibitem{muller2016interventional}
K.~M{\"u}ller, S.~Datta, M.~Ahmad, J.-H. Choi, T.~Moore, L.~Pung, C.~Niebler,
  G.~Gold, A.~Maier, and R.~Fahrig, \enquote{Interventional dual-energy
  imaging—feasibility of rapid kv-switching on a {C-arm} {CT} system,}
  Medical Physics \textbf{43}(10), 5537--5546 (2016).

\bibitem{zhu2022feasibility}
J.~Zhu, T.~Su, X.~Zhang, J.~Yang, D.~Mi, Y.~Zhang, X.~Gao, H.~Zheng, D.~Liang,
  and Y.~Ge, \enquote{Feasibility study of three-material decomposition in
  dual-energy cone-beam {CT} imaging with deep learning,} Physics in Medicine
  \& Biology \textbf{67}(14), 145012 (2022).

\bibitem{shi2020characterization}
L.~Shi, M.~Lu, N.~R. Bennett, E.~Shapiro, J.~Zhang, R.~Colbeth, J.~Star-Lack,
  and A.~S. Wang, \enquote{Characterization and potential applications of a
  dual-layer flat-panel detector,} Medical Physics \textbf{47}(8), 3332--3343
  (2020).

\bibitem{staahl2021performance}
F.~St{\aa}hl, D.~Sch{\"a}fer, A.~Omar, P.~van~de Haar, F.~van Nijnatten,
  P.~Withagen, A.~Thran, E.~Hummel, B.~Menser, {\AA}.~Holmberg \emph{et~al.},
  \enquote{Performance characterization of a prototype dual-layer cone-beam
  computed tomography system,} Medical Physics \textbf{48}(11), 6740--6754
  (2021).

\bibitem{wang2023dual}
Z.~Wang, H.~Zhou, S.~Gu, Y.~Xia, H.~Liao, Y.~Deng, and H.~Gao,
  \enquote{Dual-energy head cone-beam {CT} using a dual-layer flat-panel
  detector: Hybrid material decomposition and a feasibility study,} Medical
  Physics \textbf{50}(11), 6762--6778 (2023).

\bibitem{wang2021high}
W.~Wang, Y.~Ma, M.~Tivnan, J.~Li, G.~J. Gang, W.~Zbijewski, M.~Lu, J.~Zhang,
  J.~Star-Lack, R.~E. Colbeth \emph{et~al.}, \enquote{High-resolution
  model-based material decomposition in dual-layer flat-panel {CBCT},} Medical
  Physics \textbf{48}(10), 6375--6387 (2021).

\bibitem{zhu2023super}
J.~Zhu, T.~Su, X.~Zhang, H.~Cui, Y.~Tan, H.-R. Zheng, D.~Liang, J.~Guo, and
  Y.~Ge, \enquote{Super resolution dual-layer {CBCT} imaging with model-guided
  deep learning,} Physics in Medicine \& Biology  (2023).

\bibitem{hsieh2009tu}
J.~Hsieh, \enquote{{TU-E}-210{A}-01: Dual-energy {CT} with fast-kvp switch,}
  Medical Physics \textbf{36}(6Part24), 2749--2749 (2009).

\bibitem{tivnan2020combining}
M.~Tivnan, W.~Wang, G.~J. Gang, E.~Liapi, P.~No{\"e}l, and J.~W. Stayman,
  \enquote{Combining spectral {CT} acquisition methods for high-sensitivity
  material decomposition,} in: \emph{Medical Imaging 2020: Physics of Medical
  Imaging}, vol. 11312, 306--311 (SPIE, 2020).

\bibitem{li2020mdm}
D.~Li, D.~Zeng, S.~Li, Y.~Ge, Z.~Bian, J.~Huang, and J.~Ma,
  \enquote{{MDM-PCCT}: Multiple dynamic modulations for high-performance
  spectral {PCCT} imaging,} IEEE Transactions on Medical Imaging
  \textbf{39}(11), 3630--3642 (2020).

\bibitem{chu2013combination}
J.~Chu, W.~Cong, L.~Li, and G.~Wang, \enquote{Combination of
  current-integrating/photon-counting detector modules for spectral {CT},}
  Physics in Medicine \& Biology \textbf{58}(19), 7009 (2013).

\bibitem{holbrook2020dual}
M.~D. Holbrook, D.~P. Clark, and C.~T. Badea, \enquote{Dual source hybrid
  spectral micro-{CT} using an energy-integrating and a photon-counting
  detector,} Physics in Medicine \& Biology \textbf{65}(20), 205012 (2020).

\bibitem{yu2018dual}
L.~Yu, L.~Ren, Z.~Li, S.~Leng, and C.~H. McCollough, \enquote{Dual-source
  multienergy {CT} with triple or quadruple x-ray beams,} Journal of Medical
  Imaging \textbf{5}(3), 033502--033502 (2018).

\bibitem{tao2020multi}
S.~Tao, J.~F. Marsh, A.~Tao, G.~J. Michalak, K.~Rajendran, C.~H. McCollough,
  and S.~Leng, \enquote{Multi-energy ct imaging for large patients using
  dual-source photon-counting detector ct,} Physics in Medicine \& Biology
  \textbf{65}(17), 17NT01 (2020).

\bibitem{wang2022fast}
S.~Wang, Y.~Yang, D.~Pal, N.~J. Pelc, and A.~S. Wang, \enquote{Fast kv
  switching for improved material decomposition with photon counting x-ray
  detectors,} in: \emph{Medical Imaging 2022: Physics of Medical Imaging}, vol.
  12031, 97--102 (SPIE, 2022).

\bibitem{roessl2009cramer}
E.~Roessl and C.~Herrmann, \enquote{Cram{\'e}r--rao lower bound of basis image
  noise in multiple-energy x-ray imaging,} Physics in Medicine \& Biology
  \textbf{54}(5), 1307 (2009).

\bibitem{feldkamp1984practical}
L.~A. Feldkamp, L.~C. Davis, and J.~W. Kress, \enquote{Practical cone-beam
  algorithm,} Josa a \textbf{1}(6), 612--619 (1984).

\bibitem{tian2013projection}
X.~Tian, Z.~Yin, B.~De~Man, and E.~Samei, \enquote{Projection-based dose
  metric: accuracy testing and applications for {CT} design,} in: \emph{Medical
  Imaging 2013: Physics of Medical Imaging}, vol. 8668, 611--619 (SPIE, 2013).

\bibitem{maslowski2018acuros}
A.~Maslowski, A.~Wang, M.~Sun, T.~Wareing, I.~Davis, and J.~Star-Lack,
  \enquote{Acuros {CTS}: A fast, linear boltzmann transport equation solver for
  computed tomography scatter--part i: Core algorithms and validation,} Medical
  Physics \textbf{45}(5), 1899--1913 (2018).

\bibitem{forthmann2009adaptive}
P.~Forthmann, M.~Grass, and R.~Proksa, \enquote{Adaptive two-pass cone-beam
  artifact correction using a {FOV}-preserving two-source geometry: A
  simulation study,} Medical Physics \textbf{36}(10), 4440--4450 (2009).

\bibitem{sidky2005robust}
E.~Y. Sidky, L.~Yu, X.~Pan, Y.~Zou, and M.~Vannier, \enquote{A robust method of
  x-ray source spectrum estimation from transmission measurements: Demonstrated
  on computer simulated, scatter-free transmission data,} Journal of applied
  physics \textbf{97}(12) (2005).

\bibitem{jin2014cnr}
Y.~Jin, H.~Gao, J.~Pack, U.~Wiedmann, and B.~De~Man, \enquote{{CNR} analysis of
  dual energy technologies,} in: \emph{Proceeding of CT meeting} (2014).

\end{thebibliography}

\end{document}